\documentclass[aps,pre,showpacs]{revtex4}

\usepackage[]{graphicx}
\usepackage{bm}

\begin{document}
                                                                                
\title{Elliptical orbits in the Bloch sphere}
\author{A.~Mandilara and J.~W.~Clark}
\affiliation{Department of Physics, Washington University,
Saint Louis, MO 63130}
\author{M.~S.~Byrd}
\affiliation{Department of Physics, Southern Illinois University,
Carbondale, IL 62901}
                                                                                
\date{\today}

\begin{abstract}
As is well known, when an $SU(2)$ operation acts on a two-level system, 
its Bloch vector rotates without change of magnitude.  Considering
a system composed of {\it two} two-level systems, it is proven that 
for a class of nonlocal interactions of the two subsystems including 
$\sigma_i\otimes \sigma_j$ (with $i,j \in \{x,y,z\}$) and the Heisenberg 
interaction, the geometric description of the motion is particularly 
simple: each of the two Bloch vectors follows an elliptical orbit within 
the Bloch sphere.  The utility of this result is demonstrated in two 
applications, the first of which bears on quantum control via quantum 
interfaces.  By employing nonunitary control operations, we extend 
the idea of controllability to a set of points which are not necessarily 
connected by unitary transformations. The second application shows how 
the orbit of the coherence vector can be used to assess the entangling 
power of Heisenberg exchange interaction.
\end{abstract}
\maketitle

\section{Introduction}
The Bloch vector, or vector of coherence \cite{Alicki}, 
provides a geometric description of the density matrix 
of a spin-1/2 particle which is commonly used in nuclear
magnetic resonance.
Mathematically, the Bloch vector may be viewed as the adjoint 
representation of an $su(2)$ object in an $so(3)$ basis 
\cite{alta3}.
Extension of the notion of vector of coherence to two-spin systems
\cite{fano,quan}, and more generally to quantum spin systems of higher 
dimensions \cite{byrd}, has drawn attention in the contexts of
quantum information theory and quantum computation.  Specific
motivations include the prospects of a useful quantification
of entanglement for composite systems \cite{mahl,byrd,alta1} and
the quest for equations describing observables in quantum 
networks \cite{quan}.  

In the present work, the extension of the Bloch formalism to two
spins is used to obtain a geometric representation of the 
orbits of the vector of coherence for each spin system in the 
case that a nonlocal interaction of the form $\sigma_i\otimes\sigma_j$
is introduced.  We propose that this geometric picture will be useful
in devising schemes for control of a quantum state via quantum 
interfaces \cite{lloyd}, i.e., through the mediation of an ancillary 
system.  In this vein, we investigate the limits of control of a 
quantum state $S$, mixed or pure, given a nonlocal interaction and 
an ancilla $Q$. The simple geometric picture developed below also 
applies to another special case of nonlocal interaction, namely the 
Heisenberg exchange Hamiltonian.  As a second application of our 
formal results, we investigate the entanglement power of the 
Heisenberg interaction.

\section{Product of operator basis for a density matrix}

\subsection{One qubit}

The density matrix $\rho$ of a two-state system is a 
positive semi-definite Hermitian $2 \times 2$ 
matrix having unit trace.  It can always be given expression in terms of
the three  trace-free Pauli matrices 
$\sigma_i,~i=1,2,3 $, which are 
generators of $su(2)$, and $I/{\sqrt{2}}$ ($I$ being the unit matrix):
\begin{equation}
\rho=\frac{1}{2} I +{\bf v}{\bf \sigma}\,.
\label{onequbit}
\end{equation} 
Here $\bf v$ is the vector of coherence, whose magnitude
is bounded by $0 \leq\parallel{\bf v}\parallel\leq 
1/2$ because
$1/2\leq {\rm Tr}(\rho^2)\leq1$.  The two limiting values of 
the norm correspond to maximally mixed and pure states, respectively.  
The magnitude of the Bloch vector differs by a factor of $1/2$ from 
that of the vector of coherence, as a matter of convention.

Unitary operations rotate the Bloch vector without changing its magnitude:
$ SU(2)$ operations on the qubit correspond to $SO(3)$ operations
on the Bloch vector.  The dynamical evolution of the Bloch vector 
under non-local operations is considered in the next section.
 
\subsection{Two qubits and the correlation tensor}

In analogy to the representation (\ref{onequbit}), we adopt
the generators of ${\mit G}=SU(4)$, i.e., the elements of 
the algebra ${\mit g}=su(4)$ (together with the unit matrix),
as an orthonormal basis for the $4\times4$ density matrix of the
two-qubit system.  We employ this basis as it appears in 
Ref.~\onlinecite{alta1}, noting that it differs from the basis 
used in \cite{byrd,mahl} only in the coefficients.

The dynamical evolution of the system becomes more transparent if we 
choose basis elements of the algebra ${\mit g}= su(4)$ in accordance with 
its Cartan Decomposition ${\mit g}={\mit p}\oplus{\mit e}$ \cite{Bro,Zhang}.
The algebras ${\mit p}$ and ${\mit e}$ satisfy the commutations relations
\begin{equation}
[{\mit e},{\mit e}] \subset {\mit e}\,, \quad
[{\mit p},{\mit e}] \subset {\mit p}\,, \quad
[{\mit p},{\mit p}] \subset {\mit e}\,.
\end{equation}
The basis elements, $W_j,~j=1,\ldots,15$ of the orthogonal algebra pair 
$(e,p)$ are
\begin{equation}
{\mit e}= {\rm span} \frac{i}{2} \{\sigma_x\otimes1,\sigma_y\otimes1,
\sigma_z\otimes1, 1\otimes\sigma_x,1\otimes\sigma_y,1\otimes\sigma_z\}\,,
\label{cart1}
\end{equation}
\begin{equation}
{\mit p}= {\rm span} \frac{i}{2} \{\sigma_x\otimes\sigma_x, \sigma_x\otimes
\sigma_y,\sigma_x\otimes\sigma_z, \\ \sigma_y\otimes\sigma_x, \sigma_y\otimes
\sigma_y,\sigma_y\otimes\sigma_z,  \\
 \sigma_z\otimes\sigma_x, \sigma_z\otimes
\sigma_y,\sigma_z\otimes\sigma_z \}\,.
\label{cart2}
\end{equation}

The basis defined by Eqs.~(\ref{cart1}) and (\ref{cart2}) is used to 
expand the density matrix as  
\begin{equation}
\rho=\sum_{j=0}^{15} {\rm Tr}(\rho X_j) X_j
=\sum_{j=0}^{15}\rho_j  X_j\,, 
\end{equation}
where $X_0=I/\sqrt{4}$, $\rho_0=1/\sqrt{4}$, and $X_j= -i W_j$ ($j=1, \ldots, 15$).
In this representation, the density matrix is specified by three objects,
namely the  vectors of coherence 
${\bf r}_1$ and  ${\bf r}_2$
for the two subsystems along with the spin-spin correlation tensor $T_j^i$.
\begin{equation}
{\bf r}_1=\left( \begin{array}{c}
               \rho_1 \\ \rho_2 \\ \rho_3 \end{array}\right)\,, \qquad
{\bf r}_2=\left( \begin{array}{c}
               \rho_4 \\ \rho_5 \\ \rho_6 \end{array}\right)\,, \qquad
T_j^i=\left( \begin{array}{ccc}
                \rho_7 & \rho_8 & \rho_9 \\
              \rho_{10} & \rho_{11} & \rho_{12} \\
             \rho_{13} & \rho_{14} & \rho_{15} \end{array}\right)\,.
\end{equation}
We note that the object $ T_j^i$ has other names: Stokes
tensor \cite{Boston}, entanglement tensor \cite{mahl}, and
tensor of coherence (when combined with the coherence vectors
in one object).  Details of the properties of $T_j^i$ can 
be found in Ref.~\onlinecite{alta1}, where many prior studies
are cited.  This tensor contains information on the correlations between
the two subsystems, of both classical and quantum nature.  Necessary and 
sufficient conditions for separability of a pure state can be stated 
in terms of its properties, whereas in the case of a mixed state, 
only necessary conditions for separability can be given \cite{alta1}.

\section{Evolution Under Local and non Local Operations}

As we have seen, the Lie algebra ${\mit g}=su(4)$ possesses 
a Cartan decomposition $ {\mit g}={\mit e}\oplus {\mit p}$, 
which informs us that there exists within the Lie group ${\mit G}=SU(4)$ 
a subgroup of local operations ${\mit G}_L=SU(2)\otimes SU(2)$
generated by ${\mit e}$.  All the other operations are nonlocal 
and members of the coset space $SU(4)/SU(2)\otimes SU(2)$, which
does not form a subgroup of $SU(4)$.
It is known (see {\it Proposition 1} of Ref.~\onlinecite{Zhang}) that any 
$U\in SU(4)$ can be written as
\begin{equation}
U=k_1Ak_2
\label{decom1}
\end{equation}
with
\begin{equation}
A = \exp\left[\frac{i}{2}(c_1\sigma_{x}^{1}\sigma_{x}^{2}
+c_2\sigma_{y}^{1}\sigma_{y}^{2}+c_3\sigma_{z}^{1}\sigma_{z}^{2})\right]\,,
\label{decom2}
\end{equation}
where $k_1,~k_2 \in SU(2)\otimes SU(2)$ and $c_1,~c_2,~c_3~\in R$.
In the following, we focus on the effect of nonlocal operations generated by
a single operator among the possibilities for 
$\sigma_i\otimes\sigma_j$, where $i,j \in\{x,y,z\}$.  
(This consideration includes the special case in which two of the
parameters $c_1$, $c_2$, and $c_3$ in the decomposition 
(\ref{decom1})-(\ref{decom2}) are zero.)  
Such nonlocal operations will be called one-dimensional.

\subsection{Local operations}

Local operations are operations $g\in SU(2)\otimes SU(2)$ 
generated by the elements of ${\mit e}$.  From the commutation 
relations $[{\mit e},{\mit e}] \subset {\mit e}$ and
$[{\mit p},{\mit e}] \subset {\mit p}$ it is clear that 
the elements of the vectors ${\bf r}_i$ and tensor $(T_i^j)$ 
do not mix and do not affect one another.  Under local operations, 
the vectors behave just like ordinary Cartesian vectors.
In particular, a vector of coherence is rotated about 
some vector $\hat n$ as illustrated in Fig.~1, i.e.,
\begin{equation}
(r')_{1}^{i}=R^{i}_{j}r_1^j\,, 
\qquad (r')_{2}^{i}=R^{i}_{j}r_2^j\,.
\end{equation} 
On the other hand, the correlation tensor transforms like
a mixed Cartesian tensor,
\begin{equation}
(T')^{i}_{j}=R^{i}_{m}(R')^{n}_{j} T^{m}_{n}\,.
\end{equation}
The magnitude of each object remains invariant under local 
operations.  In addition, there exist fifteen more invariants which
can be constructed from the vectors and the tensor \cite{makhl}.  
 
\begin{figure}
\includegraphics[width=12cm]{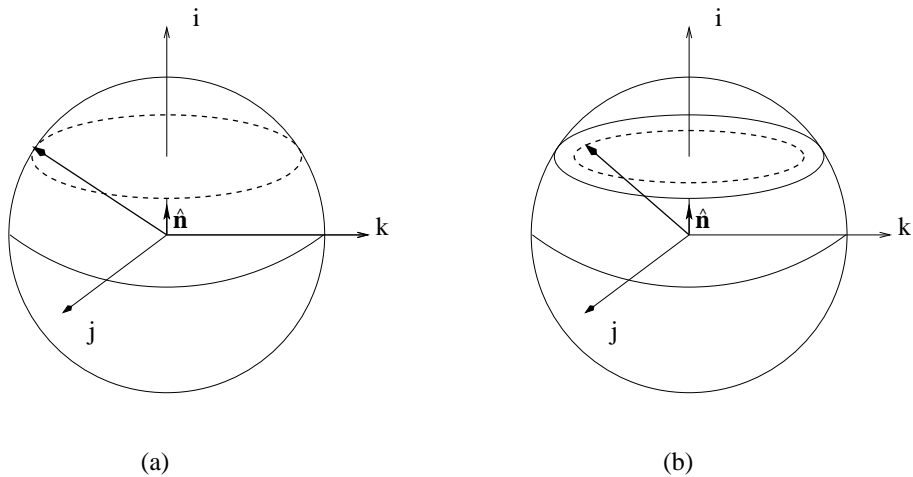}
\caption{Local operations on the two spin subsystems produce
a rotation of the corresponding vectors of coherence 
around some direction 
$\hat{{\bf n}}$. 
The effect is the same for both pure states (a) and mixed states (b).}
\end{figure}

\subsection{One-dimensional nonlocal operations }

The nonlocal operations in the coset space $SU(4)/SU(2)\otimes SU(2)$
require, in their construction, exponentiation of at least one of the
elements of ${\mit p}$.  Hence, under these operations 
the elements of the tensor and vectors of coherence are mixed, 
due to the commutation relations $[{\mit p},{\mit e}] \subset {\mit p}$
and $[{\mit p},{\mit p}] \subset {\mit e}$.  We shall establish
that the one-dimensional nonlocal operations generated by the 
chosen interaction
$\sigma_i\otimes\sigma_j$ give rise to elliptical orbits for 
the vectors of coherence of the subsystems.  The characteristics 
of these elliptic paths depend on the indices $i$ and $j$, on 
the initial states of the subsystems, and on the degree of 
correlations between 
them.  These orbits can be described by 
non-unitary transformations on each of the individual subsystems 
when one traces over the other's degrees of freedom.

Accordingly, we take the interaction Hamiltonian between the two 
spins to be $H_I=\sigma_i\otimes\sigma_j/2$, and, for
reasons of simplicity, we suppose that the internal Hamiltonians 
for the two spins may be ignored.  Assuming that the duration of
the interaction is $\phi$, and appealing to (i) the commutation 
relations as summarized in Ref.~\cite{Zhang} and (ii) the identity
\begin{equation}
\exp\left[-i(\phi/2)\sigma_i\otimes\sigma_j\right] =
\cos(\phi/2)I-i\sin(\phi/2)\sigma_i\otimes\sigma_j\,, 
\end{equation}
we can make the following observations:
\begin{enumerate}
\item[(i)] The components $r^{i}_1$ and $r^{j}_2$ of the vectors
of coherence remain unaffected; hence the vectors are confined
to planes perpendicular to the $i$-axis and $j$-axis respectively.
\item[(ii)] Of the nine elements of the correlation tensor $T^{k}_{l}$, 
only four experience changes.
 The five that are unchanged under the 
action of $\sigma_i\otimes \sigma_j$ are $T^i_j$ and $T^k_l$ with 
$k\neq i$ and $l\neq j$.
\item[(iii)] The vectors $r_1^m+T_j^m$ and $r_2^m+T_m^i$ are rotated, without change
of magnitude, through an angle $\phi$ about the $i$ and $j$ axes,
respectively. (Here $m$ ranges freely over $\{x,y,z\}$).
\item[(iv)] More explicitly, the components of the vectors transform
according to
\begin{equation}
\begin{array}{l}
r_{1}^{i}\rightarrow  (r')_{1}^{i}=r_{1}^{i}\,, \nonumber \\
r_{1}^{k}\rightarrow  (r')_{1}^{k}=r_1^k \cos \phi- T^{l}_{j}\sin \phi 
\,,\nonumber \\
r_{1}^{l}\rightarrow  (r')_{1}^{l}= T^{k}_{j}\sin \phi + r_{1}^{l}\cos \phi \,,
\end{array}
\begin{array}{l}
r_{2}^{j}\rightarrow  (r')_{2}^{j}=r_{2}^{j}\,,\nonumber  \\
r_{2}^{m}\rightarrow  (r')_{2}^{m}=r_{2}^{m}\cos \phi- T^{i}_{n}\sin \phi\,,\nonumber\\
r_{2}^{n}\rightarrow  (r')_{2}^{n}= T^{i}_{m}\sin \phi +r_{2}^{n}\cos \phi \,,
\end{array}
\end{equation}
and the components of the tensor of coherence, according to
\begin{equation}
\begin{array}{l}
T_{j}^{i}\rightarrow  (T')_{j}^{i}=T_{j}^{i}\,,\nonumber \\
T_{j}^{k}\rightarrow  (T')_{j}^{k}=T^{k}_{j}\cos \phi-r^{l}_{1}\sin \phi \,,\nonumber\\
T_{j}^{l}\rightarrow  (T')_{j}^{l} =r^{k}_{1}\sin \phi + T_{j}^{l} \cos \phi \,,
\end{array},~~
\begin{array}{l}
 T_{j}^{i}\rightarrow  (T')_{j}^{i}=T_{j}^{i}\,,\nonumber \\
T_{m}^{i}\rightarrow  (T')_{m}^{i}=T^{i}_{m}\cos \phi -r^{n}_{2}\sin \phi \,,\nonumber\\
T_{n}^{i}\rightarrow  (T')_{n}^{i} =r^{m}_{2}\sin \phi + T_{n}^{i}\cos \phi \,,
\end{array}
\end{equation}
with no change in the tensor's other elements.
The ordered sets of indices $(i,l,k)$ and $(j,m,n)$ belong to 
$\{(x,y,z),(y,z,x),(z,x,y)\} $.
\end{enumerate}
Given this behavior, it is not difficult to show that {\it ${\bf r}_1(\phi)$ and 
${\bf r}_2(\phi)_2$ follow elliptical orbits}.  Since the 1,2 labeling is 
arbitrary, it suffices to demonstrate this property for the the 
vector ${\bf r}_1(\phi)$. 
\smallskip

\noindent
{\it Proof.}
Referring to Fig.~2(a), the coordinates for a vector $\bf s$ tracing 
an ellipse in the $x-y$ plane, with principal axes $a$ and $b$ rotated 
by an angle $\psi$, are 
\begin{equation}
\begin{array}{l}
s_x(\phi)=  a ~{\rm cos}\phi~{\rm cos}\psi+ b~{\rm sin}\phi~{\rm sin}\psi\,,\nonumber\\
s_y(\phi)= - a~ {\rm cos}\phi~{\rm sin}\psi+ b~{\rm sin}\phi~{\rm cos}\psi \,.
\end{array}
\end{equation}
The angle $\phi$ is zero when the vector $\bf s$ is aligned with 
the principal axis $a$.

The coordinates of the vector of coherence ${\bf r}_1$ moving in the
$k-l$ plane are given by
\begin{equation}
\begin{array}{l}
r_1^k(\phi')=r_1^k(0)\cos \phi' - T_j^l(0) \sin \phi'\,,\nonumber \\
r_1^l(\phi')=T_j^k(0)\sin \phi'+r_1^l(0) \cos \phi' \,.
\end{array}
\end{equation}
Of course, for the vector of coherence, $\phi' = 0 $ does not in general
correspond to the principal axis $a$ (see Fig.~2(a)).  In fact,
$\phi'=\phi+\chi$, and the coordinates of ${\bf r}_1 $ can be rewritten
as follows in terms of the phase difference $\chi$:
\begin{equation}
\begin{array}{l}
r_1^k(\phi)=(r_1^k(0)\cos\chi- T_j^l(0)\sin\chi)\cos\phi+
        (-T_j^l(0)\cos\chi-r_1^k(0)\sin\chi)\sin\phi \,, \nonumber\\
 r_1^l(\phi)=(T_j^k(0)\sin\chi+r_1^l(0)\cos\chi) \cos\phi
       +(-r_1^l(0)\sin\chi+T_j^k(0)\cos\chi)\sin\phi \,.
\end{array}
\end{equation}
Comparison of the two sets of coordinates $\{s_x(\phi),s_y(\phi)\}$ 
and $\{r_1^k(\phi),r_1^l(\phi) \}$ shows that a match can always
be made, such that the parameters $a$, $b$, $\psi$, and $\chi$
can be determined by solving the system of equations
\begin{equation}
\begin{array}{c}
a\cos\psi= r_1^k(0)\cos\chi- T_j^l(0)\sin\chi \,, \nonumber \\
b\sin\psi=-T_j^l(0)\cos\chi-r_1^k(0)\sin\chi \,, \nonumber  \\ 
a\sin\psi=-T_j^k(0)\sin\chi-r_1^l(0)\cos\chi \,, \nonumber  \\ 
b\cos\psi=-r_1^l(0)\sin\chi+T_j^k(0)\cos\chi\,. 
\label{system}
\end{array} 
\end{equation}
This completes the proof.
It is important to note that the shape of the ellipse depends
explicitly on the spin-spin correlation tensor.

\smallskip

\begin{figure}
\includegraphics[width=12cm]{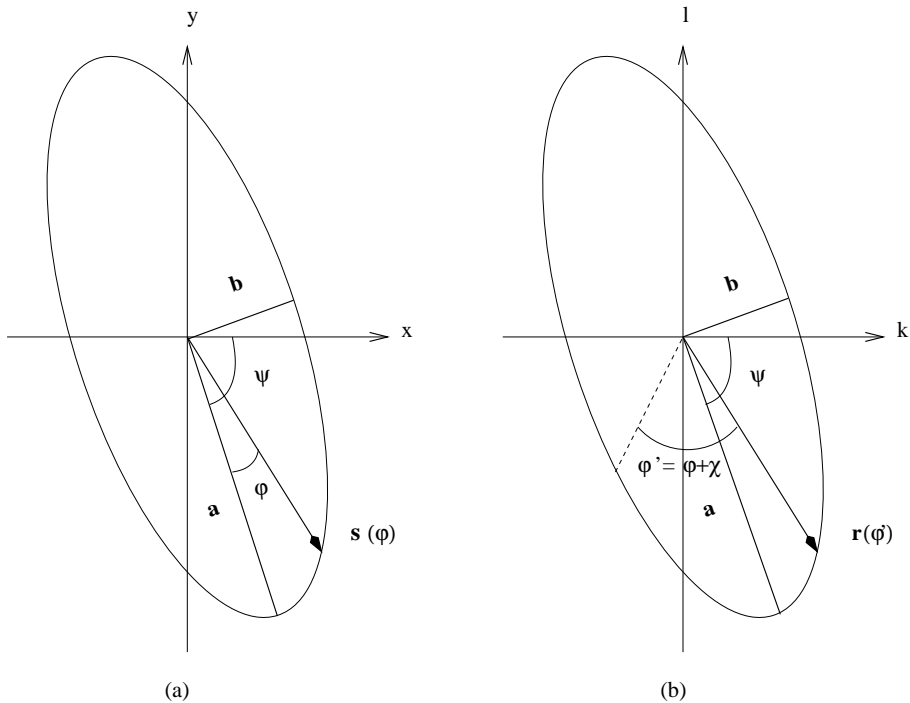}
\caption{(a) The vector $\bf s$ describes an ellipse with principal 
axes $a$ and $b$ rotated by angle $\psi$ with respect to the $y$-axis. 
The angle $\phi$, interpreted as the duration of a group operation, is 
measured relative to the $a$ principal axis.  (b) The vector of coherence 
$\bf r$ corresponding to one of the spins of the two-spin system moves 
on an ellipse on the $k-l$ plane, with the angle $\phi'$ measured relative to 
the original direction of $\bf r$.}
\end{figure}

Solving Eqs.~(\ref{system}) for the angle $\chi$, we find 
\begin{equation}
\tan(2\chi)= \frac{2\left[r^{k}_{1}(0)T^{l}_{j}(0)-r^{l}_{1}(0)T^{k}_{j}(0))\right]}
{-(r^{l}_{1}(0))^2+(T^{k}_{j}(0))^2 -(r^{k}_{1}(0))^2+(T^{l}_{j}(0))^2}\,,
\end{equation}
which specifies the initial orientation of the coherence vector ${\bf r}_1$
with respect to the principal axis $a$.
Suppose now the two-spin system is initially in a {\it product state}. 
For this case it is easy to prove these corollaries to our principal result:
\begin{itemize}
\item[(1)] The phase difference $\chi$ is zero.  This means that the 
initial positions of both coherence vectors lie on the $a$ 
principal axis (as in Fig.~3(a)). 
It follows that the linear entropy of the state of each of the 
subsystems (defined by $1 - {\rm Tr}\,\rho^2$) can only decrease, 
showing it is possible to increase the entanglement of the system 
with this interaction.  (This will depend on initial conditions.  
See Section \ref{sec:ent-heis}.)  
\item[(2)] The length of the semi-minor axis of the ellipse followed by subsystem
1 is given by $ b_1=|r_2^j(0)|[(r_1^l(0))^2+(r_1^k(0))^2]^{1/2}$
(and likewise for subsystem 2 with $1 \rightarrow 2$ 
and $\{j,k,l\} \rightarrow \{i,n,m\}$).
It follows that for an initially pure state and the assumed
single interaction $\sigma_i \otimes \sigma_j$, the
maximum attainable entanglement is achieved at $\phi$ values of
$\pi/2$ and $3 \pi/2$. 
\end{itemize}

For the case of a initial state that is not pure but still
separable, the phase difference $\chi$ does not vanish,
in general (see figure~3(b)). 
Accordingly, the implied dynamical behavior of a classically 
correlated system distinguishes it from an uncorrelated system, 
but not from a system experiencing quantum entanglement.
Moreover, the linear entropy of each subsystem can either
increase or decrease,
showing it is possible to increase or decrease the amount of 
entanglement in the system.

\begin{figure}
\includegraphics[width=12cm]{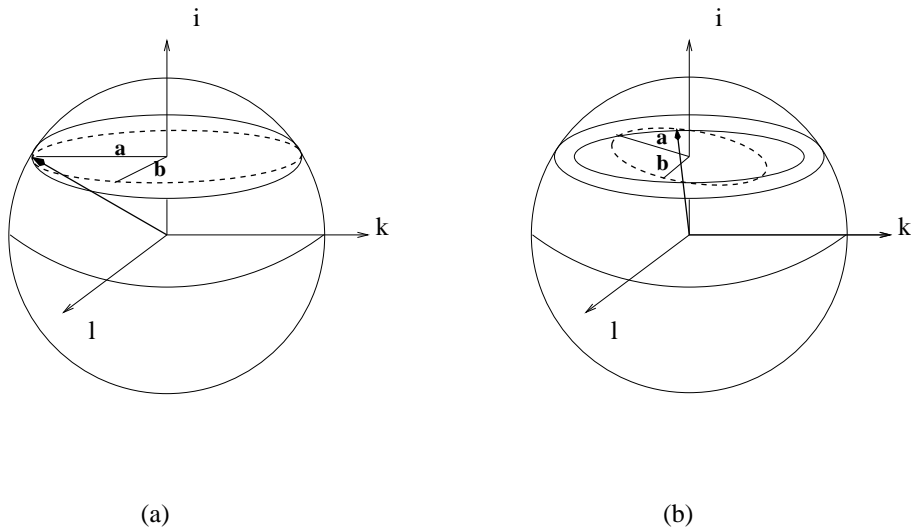}
\caption{The initial position of the vector of coherence of subsystem
1 or 2 is shown, together with its time evolution under a one-dimensional
nonlocal interaction (dashed line).  If the initial state of the two-spin
system is a product state, then the initial position is on the $a$ principal 
axis of the elliptical path, as in (a).  In general this agreement no longer
occurs if the subsystems are initially correlated, either classically or 
quantum mechanically, as in (b).} 
\end{figure}

\subsection{General nonlocal operations}

From {\it Proposition 1} of Ref.~\onlinecite{Zhang}, any nonlocal operation 
can be decomposed as a product of two local operations and an operation of the form
\begin{equation}
A= \exp\left[ \frac{i}{2}(c_1\sigma_{x}^{1}\sigma_{x}^{2}
+c_2\sigma_{y}^{1}\sigma_{y}^{2}+c_3\sigma_{z}^{1}\sigma_{z}^{2})\right] \,.
\label{decom3}
\end{equation}
The operators $\{Y_i\}=\{i\sigma_{x}^{1}\sigma_{x}^{2}/2,
i\sigma_{y}^{1}\sigma_{y}^{2}/2,i\sigma_{z}^{1}\sigma_{z}^{2}/2 \}$ span
a maximal Abelian subalgebra of ${\mit P}$, and the relations
\begin{equation}
[Y_i,Y_j]=0\,, \qquad 
[Y_i,Y_j]_+ = -i|\epsilon_{ijk}|Y_k -\frac{1}{2}\delta_{ij}
\end{equation}
hold, where $[\cdot,\cdot]_+$ denotes the anticommutator.  
Consequently, $A$ of Eq.~(\ref{decom3}) can be written
in product form,
\begin{equation}
A =\exp\left[\frac{i}{2}(c_1\sigma_{x}^{1}\sigma_{x}^{2})\right]
\exp\left[\frac{i}{2}(c_2\sigma_{y}^{1}\sigma_{y}^{2})\right]
\exp\left[\frac{i}{2}(c_3\sigma_{z}^{1}\sigma_{z}^{2})\right]\,.
\label{prodform}
\end{equation}
The property (\ref{prodform}) tells us that {\it any nonlocal operation 
can be decomposed into a sequence of operations effecting a succession
of circular and elliptic paths in Bloch space}.
This result facilitates the calculation of the final states of
the subsystems, but gives only limited insight into the geometric
characteristics of the coherence vectors' time orbits.  For 
all $c_i$ distinct, two general observations can be made:
\begin{itemize}
\item[(1)] The motion of the vectors of coherence is no longer restricted
to a plane, since there is no linear combination of 
$\{\sigma_x\otimes 1, \sigma_y\otimes 1, \sigma_z\otimes 1\}$ 
or $\{1\otimes\sigma_x,1\otimes\sigma_y,1\otimes\sigma_z\}$
that is invariant under the action of $A$.
\item[(2)] Characteristics of the trajectories such as
periodicity depend in detail on the parameters $c_1$, $c_2$, and $c_3$.  
A trajectory is periodic only if $c_2/c_1$ and $c_3/c_1$ are both rational 
numbers.  We note also that the set of parameters $\{c_1,c_2,c_3\}$ has 
been used  to determine the equivalence classes 
of nonlocal interactions \cite{Zhang} as well as the invariants of the nonlocal 
interactions \cite{makhl}. 
\end{itemize}
\subsection{Special case of the Heisenberg Hamiltonian}
The Heisenberg exchange Hamiltonian, corresponding to $c_1=c_2=c_3=-c/2$, 
is not included in our general observations on nonlocal interactions 
(made for all $c_i$ distinct), but like the one-dimensional Hamiltonians, 
it admits a simple geometric picture.  This interaction is the primary 
two-qubit interaction in several experimental proposals for quantum-dot qubits  
\cite{Loss:98, Kane:98, Vrijen:00}.  It can also be used for universal 
quantum computing on encoded qubits of several types 
\cite{Bacon:00, Kempe:01, DiVincenzo:00a, Lidar:02, Wu:02, Byrd:02}. 
For these reasons, it warrants special attention.

Introducing the time parameter $\phi$, the operator $A$ of 
Eq.~(\ref{prodform}) now takes the form 
\begin{eqnarray*}
A(\phi) &=& \exp[-i(c\phi/2)(\sigma_x\otimes \sigma_x 
                             + \sigma_y\otimes \sigma_y 
                             + \sigma_z\otimes \sigma_z)] \nonumber \\
  &=& \left[\cos^3(c\phi/2) -i\sin^3(c\phi/2)\right]I\otimes I \nonumber \\
   && -(i/2)e^{ic\phi/2}\sin(c\phi)(\sigma_x\otimes \sigma_x 
                             + \sigma_y\otimes \sigma_y 
                             + \sigma_z\otimes \sigma_z ). 
\end{eqnarray*}
The time development of the density matrix under the operator
$A$ is given $\rho(\phi)= A(\phi)\rho(0)A^{\dagger}(\phi)$
and the corresponding coherence vectors change according to 
\begin{equation}
\label{eq:ipart}
r_1^i(\phi) = \frac{1}{2}[r^i_1(0)+r^i_2(0)+(r_1^i(0)-r_2^i(0))\cos(2c\phi)
                     +(T_{k}^{j}(0)-T_{j}^{k}(0))\sin(2c\phi)] \,,
\end{equation}
where $i,j,k =1,2,3$ and cyclic permutations are implied.
Similarly, for the coherence tensor we have
\begin{equation}
\label{eq:tpart}
T_j^i(\phi) = \frac{1}{2}[T^{i}_{j}(0)+T^{j}_{i}(0)
                     +(T^{i}_{j}(0)-T^{j}_{i}(0)) \cos(2c\phi)
                 +(r_1^k(0)-r_2^k(0)) \sin(2c\phi)] \,.
\end{equation}
The elements of 
${\bf r}_2(\phi)$ 
are found by symmetry $1\leftrightarrow 2$. 
The quantities $r_1^i+r_2^i$, $T^{i}_{j}+T^{j}_{i}$, and $T^{i}_{i}$
are unchanged by the operation, and the form of the one-parameter
set that describes the time-evolving coherence vector is
\begin{equation}
{\bf r}_1(\phi) = {\bf R} + {\bf S}\cos(2c\phi) + {\bf V}\sin(2c\phi)\,,
\end{equation}
where ${\bf R}={\bf r}_1(0)+{\bf r}_2(0)$, 
${\bf S} = {\bf r_1}(0)-{\bf r}_2(0)$, and 
\begin{equation}
{\bf V} = \left(\begin{array}{c} T^{3}_{2}(0)-T^{2}_{3}(0) \\ 
                            T^{1}_{3}(0)-T^{3}_{1}(0) \\
	                    T^{2}_{1}(0)-T^{1}_{2}(0) \end{array}\right)\,.
\end{equation}
Clearly the vector traces out an ellipse lying in the plane spanned by 
${\bf S}$ and ${\bf V}$, defined by 
${\bf S}\times{\bf V}$, and 
passing through the point ${\bf R}$.

\section{Applications}
We shall now illustrate some of the results of Section III with 
two examples.  The first provides a controllability result 
for nonlocal unitary interactions and the second demonstrates how the orbit 
of the coherence vector can be used to describe the entangling power of the 
Heisenberg exchange interaction.  
\subsection{Quantum control via quantum controllers and one-dimensional 
nonlocal interactions}
Let us now consider the implications of the findings of the preceding
sections for the problem of quantum control.  To this end, we 
adopt the nomenclature of Lloyd \cite{lloyd} 
and identify spin 1 with the system $S$ whose quantum state we wish 
to control, and spin 2 with the quantum controller or interface $Q$.
It is assumed that (i) only one interaction Hamiltonian $H_I$ is
in play between $S$ and $Q$ and (i) system $Q$ is completely 
controllable via control Hamiltonians 
$\{H_Q^{m}\}=\{1\otimes \sigma_x, 1\otimes \sigma_y,1\otimes \sigma_z\} $
that span the $su(2)$ algebra. The initial state of the bipartite
system is taken to be a product state in the ensuing analysis.

Suppose that the interaction Hamiltonian is nonlocal, but takes the
one-dimensional form $H_I=\sigma_i\otimes \sigma_j $.  Then the set 
$\{\{H_Q^{m}\},H_I\}$ $=\{1\otimes \sigma_x, 1\otimes \sigma_y,
1\otimes \sigma_z, \sigma_i\otimes \sigma_x,
\sigma_i\otimes \sigma_y,\sigma_i\otimes \sigma_z\} $
comprises a closed six-element subalgebra ${\mit G}_6$ of ${\mit G} $.
Given this set of operations, the vector of coherence ${\bf r}_S$ 
of system $S$ remains in the plane perpendicular to the $i$-axis.

It has been established in Sec.~III that when $H_I=\sigma_i\otimes \sigma_j $ 
is the only element of the algebra $su(4)$ affecting the two-spin system,
the vectors of coherence ${\bf r}_1$ and ${\bf r}_2$ are constrained 
to move in elliptical orbits.  Now, with the six-element subalgebra ${\mit G}_6$
available to the two-spin system $S+Q$, the reachable set of the system $S$ 
is enlarged to an {\it elliptical disk} (see Fig.~4).  The principle axis of the
disk coincides with the initial coherence vector ${\bf r}_S(0)$ of
the $S$ system, while the length of its semiminor axis is given by 
$b=[(r_S^k(0))^2+(r_S^l(0))^2]^{1/2}|{\bf r}_Q(0)|$, where ${\bf r}_Q(0)$ is
the initial coherence vector of system $Q$. 
\smallskip

\noindent
{\it Proof.} First, if one implements the two-step sequence of a local
operation $\in 1\otimes su(2)$ on system $Q$ followed by the nonlocal
operation $H_I=\sigma_i\otimes \sigma_j$ on $S+Q$, the orbit 
of ${\bf r}_S$ is necessarily an ellipse whose $a$ principle
axis lies along the initial coherence vector ${\bf r}_S(0)$ 
and whose semimajor axis $b$ is restricted by 
$0\le b\le [(r_S^k(0))^2+(r_S^j(0))^2]^{1/2}|{\bf r}_Q(0)|$.  
Hence the set reachable by this two-step procedure is the elliptic
disk in question.  Second, using the Baker-Campbell-Hausdorff
formula one can show that all the elements of the six-element
subalgebra ${\mit G}_6$ can be constructed by this two step sequence,
so their reachable sets are the same.

From this result we infer that {\it the entropy of 
system $S$ cannot be decreased by intervention of the quantum 
interface $Q$ if the interaction Hamiltonian is limited to 
the form $H_I=\sigma_i\otimes \sigma_j$}.  Noting that 
$|{\bf r}_Q|\le 1/2$, it follows that 
$a\ge b$,
where $a$ and $b$ 
are respectively the magnitudes of the semimajor and semiminor 
axes of the elliptical reachable set.  Furthermore, it is 
seen that the systems $S$ and $Q$ become maximally entangled
if the initial state of the system $S$ is situated on the equatorial 
plane perpendicular to $i$-axis.
\begin{figure}
\includegraphics[width=12cm]{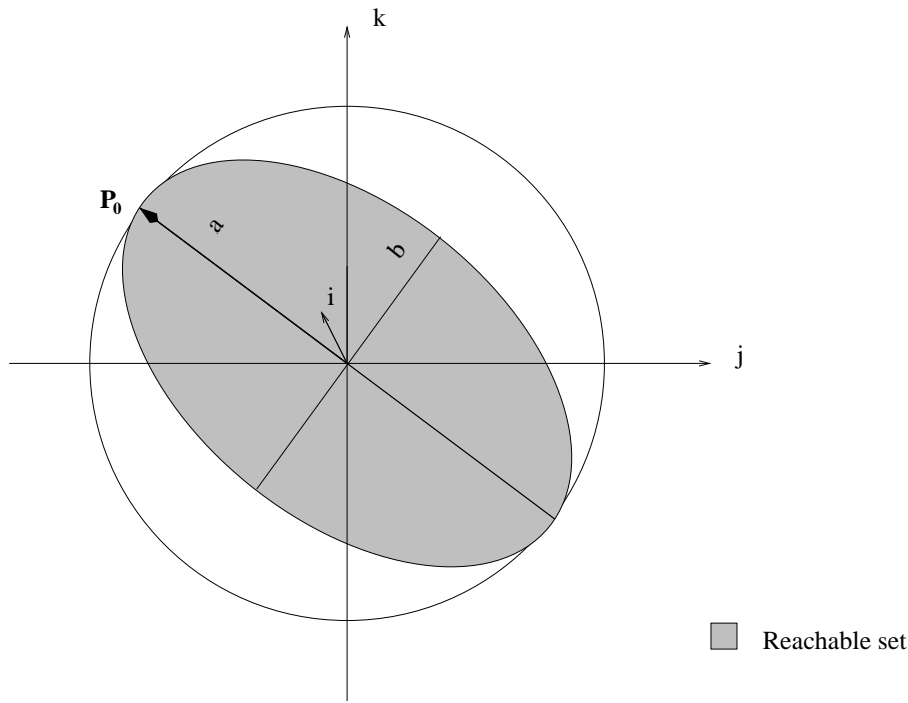}
\caption{The gray area is the set of reachable states for
the system $S$ if one has full control of the controller $Q$ and
the interaction $\sigma_i\otimes\sigma_j$ is available.  This
elliptical disk is characterized by a semimajor axis coincident
with the initial vector of coherence for $S$
and a semiminor axis with $b= [(r_S^k(0))^2+(r_S^j(0))^2]^{1/2}|{\bf r}_Q(0)|$.}
\end{figure}

\subsection{Entanglement power of Heisenberg interaction}
\label{sec:ent-heis}
Upon examining Eq.~(\ref{eq:ipart}), we see that the maximum entanglement,
realized in a maximally entangled pure state, can be achieved if 
${\bf {r}}_1(0) = - {\bf {r}}_2(0)$, $|{\bf r}_1|=1/2$, and 
$c\phi = \pm\pi/4$.  Otherwise, the state is not perfectly entangled 
since the linear entropy 1-Tr($\rho^2$) is not minimized.  This
conclusion agrees with the result of Zhang {\it et al.} \cite{Zhang} 
that the only perfect entanglers that can be achieved with the
Heisenberg Hamiltonian are the square-root of swap and its inverse.

However, suppose that the initial state of the two-spin system
is represented by
$$
\rho(0) = \frac{1}{4}(I + \sigma_z)\otimes (I + \sigma_z) \,,
$$
which is a pure-state density matrix for
which $r_1^z = 1/2 = r_2^z$ 
and $T^{z}_{z} = 1/2$, all other elements of the coherence vectors
and coherence tensor being zero.  Then 
$$
r_1^{x}(\phi) = r_1^z(0)\cos^2(c\phi)+r_2^z(0)\sin^2(c\phi)\,,
$$
while all other components of ${\bf r}_1$ and ${\bf r}_2$ vanish at time $\phi$, 
and all other $r_1^\alpha(0) =0$. 
In this case the ellipse collapses to a line and the coherence vector 
simply oscillates between two values along that line.  
The only element of the correlation tensor that changes is 
$$
T^{1}_{2} = \frac{1}{2}\sin(2c\phi)(r_1^z(0)-r_2^z(0))\,,
$$
which vanishes for an initial tensor product of pure states for which the 
subsystems are polarized in the $+z$ direction.  Therefore one cannot 
create maximally entangled states with these initial conditions.  

\section{Conclusions}
In this paper we have developed a geometric representation for the 
orbits of the coherence vectors of a two-qubit system.  In various 
circumstances we have shown that their evolution is described 
by elliptical orbits lying within the surface of the Bloch sphere.  
Importantly, every two-qubit unitary operation can be expressed as a 
combination of one of the evolutions we have considered, together
with ``pre'' and ``post'' local single-qubit rotations.
We anticipate that this geometric picture will be helpful in devising 
schemes for control of a quantum state via quantum interfaces, and 
we have obtained a controllability result appropriate for such applications.  
Given the utility of the coherence-vector picture for modeling 
quantum systems and describing their entanglement, 
further studies along similar lines may be fruitful.  Such
work could include analysis of the orbits of higher-dimensional 
quantum states, as well as consideration of the effects of measurement 
operations on controllability.

\section*{Acknowledgments}
This research was supported by the U.~S.\ National Science Foundation
under Grant No.~PHY-0140316 (JWC and AM) and by the Nipher Fund.

\end{document}